\documentclass{aa}  

\usepackage{graphicx}
\usepackage{amssymb,amsmath}
\usepackage{hyperref}			
\usepackage{lmodern}
\usepackage{mathtools}
\usepackage[varg]{txfonts}

\makeatletter
\renewcommand*\aa@pageof{, page \thepage{} of \pageref*{LastPage}}
\makeatother

\begin{document}
\title{Low thermal conductivity of the super-fast rotator (499998) 2011 PT}

   \author{Marco Fenucci\inst{1}
          \and
          Bojan Novakovi\'c\inst{1}
          \and
          David Vokrouhlick\'y\inst{2}
          \and
          Robert J. Weryk\inst{3}
          }
   \authorrunning{Fenucci et. al.}
   \institute{Department of Astronomy, Faculty of Mathematics, University of Belgrade,
             Studentski trg 16, 11000 Belgrade, Serbia\\
               \email{marco\_fenucci@matf.bg.ac.rs}
               \and
               Institute of Astronomy, Charles University, V Holešovičkách 2, CZ-180 00 Prague 8, Czech Republic
               \and
               Institute for Astronomy, University of Hawaii, Honolulu HI, 96822, USA
             }
   \date{Received --- / Accepted ---}


  \abstract
     {Asteroids up to a few tens of meters in diameter may spin very fast, completing an
     entire rotation in a period of few minutes. These small and fast rotating
     bodies are thought to be monolithic objects, since the weak gravitational force
     due to their small size is not strong enough to counteract the large centripetal
     force caused by the fast rotation. This argument makes the rubble-pile structure not feasible for
     such objects. Additionally, it is not clear whether the fast spin prevents dust and small particles (regolith)
     to be kept on their surface.}
     {We aim to develop a model to constrain the thermal conductivity of the surface of the small, fast-rotating near-Earth asteroids. This model may suggest whether the presence of regolith is likely or not.}
     {Our approach is based on the comparison between the measured
      Yarkovsky drift and a predicted value using a theoretical model, which depends on the orbital, physical and thermal parameters of the object. The necessary parameters are either deduced from statistical distribution derived for near-Earth asteroids population or determined from observations with associated uncertainty. With that information available, we perform Monte Carlo simulations, producing a probability density distribution for the thermal conductivity.}
     {Applying our model to the super-fast rotator asteroid (499998) 2011 PT, we find that the measured Yarkovsky drift can be achieved only if the thermal conductivity $K$ of the surface is low. The resulting probability density function for the conductivity is bimodal, with two most likely values being around 0.0001 and 0.005 W m$^{-1}$ K$^{-1}$. Based on this, we find that the probability of $K$ being smaller than 0.1 W m$^{-1}$ K$^{-1}$ is at least 95 per cent. This low thermal conductivity could be a clue that the surface of 2011 PT is covered with a thermal insulating layer, composed by a regolith-like material similar to lunar dust.
     }
     {}



   \maketitle

\section{Introduction}

Knowledge of the surface properties of asteroids provides insights into the nature 
of the surface materials and structures, which are, in turn, fundamental for different reasons.
These properties are, for instance, important in modeling the formation of regolith-like materials \citep{delbo-etal_2014}, the weakening and degradation of boulders
on asteroids due to thermal shocks \citep{2017Icar..294..247M}, and 
space weathering processes \citep{2015aste.book..597B}. 
Moreover, surface properties are also crucial 
for planning of spacecraft interaction with the surface of an asteroid, including deflection missions \citep{2016Icar..269...50B}. 

The surface thermal inertia informs us about porosity and cohesion, with low values 
being indicative of the presence of a thermal insulting layer \citep[see][and references therein]{2020A&A...638A..84A}, consistent with a dusty, 
porous regolith, while high values are specific for rocky-like materials. Thermal inertia depends on density $\rho$, heat capacity $C$ and thermal conductivity $K$. However, while density and heat capacity typically vary within a factor of a few, uncertainty in thermal conductivity spans over a range of more than four orders of magnitude \citep{delbo-etal_2015}. Therefore, in this respect,
the thermal conductivity is the most important parameter to constrain.

Before in-situ exploration of asteroids, it was thought that
only very large objects could be covered by regolith \citep[see][and references therein]{2015aste.book..767M}. 
However, the visit of the Galileo and NEAR-Shoemaker NASA missions to asteroids (243) Ida, (433) Eros, and (951) Gaspra
revealed that even smaller asteroids can actually retain regolith-like materials
on their surfaces. Furthermore, with the visit of the JAXA Hayabusa mission to asteroid (25143) 
Itokawa in 2005, it became evident from its in-situ observations and measurements
that, despite the low gravitational environment, even sub-kilometer bodies can retain gravel particles, boulders or regolith on their surfaces \citep{miyamoto-etal_2007}.
More recently, data collected in-situ on asteroids (101955) Bennu and (162173) Ryugu by the OSIRIS-REx 
and the Hayabusa~2 missions, respectively, did not reveal fine dust on their surfaces, but still
showed the presence of grain particles and boulders of different sizes \citep{lauretta-etal_2019, michikami-etal_2019,morota-etal_2020, sugita-etal_2020, susorney-etal_2020}.

These findings changed our perspective on the ability of kilometer and sub-kilometer sized asteroids
to preserve regolith-like materials on their surfaces. This new perspective raises the question if even smaller
objects, from a few tens to a few hundreds of meters, can also be covered by a dust layer. The extremely low-gravity environment on such bodies would suggest that a dust layer is not likely. If we further restrict our considerations to small super-fast rotators, such a possibility seems even less probable.

\begin{figure}[!ht]
   \centering
   \includegraphics[width=0.45\textwidth]{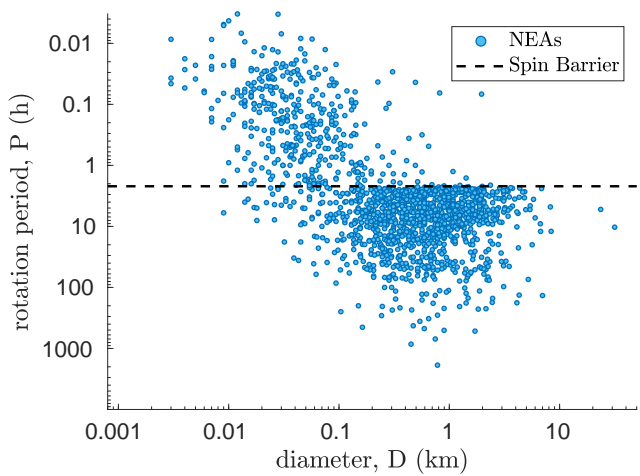}
   \caption{Measured spin periods for near-Earth objects as of September 2020 and taken from the Asteroid Lightcurve Database
   \citep[see][]{warner-etal_2009}.}
   \label{fig:NEAspinPeriods}
\end{figure}

\citet{pravec-harris_2000} found that objects larger than
about 150 meters very rarely have rotation period smaller than about 2.2 hr. This barrier can be
explained by the fact that, for diameters between about 150 meters and 10 kilometers,  every Solar 
System body has a rubble-pile internal structure \citep{walsh_2018}, which is not gravitationally
strong enough to counteract the centripetal force due to the fast rotation, thus causing a break-up of the
object. On the other hand, there is no such limit for the spin period of smaller
objects, so that they are allowed (and often seen) to rotate very fast, even with periods of few minutes. 
Asteroids smaller than 150 meters in diameter are therefore considered to be mostly monolithic objects
\citep[see, e.g.,][]{pravec-harris_2000, whiteley-etal_2002, polishook_2013}, with strong internal forces that hold the body intact 
even at very fast spin rates.
Figure~\ref{fig:NEAspinPeriods} shows the measurements of the spin period of near-Earth
objects (NEOs), as a function of the diameter of the body, taken from the Asteroid Lightcurve
Database\footnote{\url{http://alcdef.org/}} (LCDB) developed by \citet{warner-etal_2009}. We can clearly 
see the spin barrier for diameters larger than 150 meters, while going down to a few tens of
meters the spin periods can be as short as few to few tens of minutes
(possibly even shorter, but typical observation scenarios would not detect them). 
Such fast spin rates, coupled with the very low gravitational environment due to the small size,
are expected to cause the ejection of material on the surface,
thus preventing regolith grains to be retained.

Determining values of thermal parameters for asteroids is a challenge, since direct measurements from Earth 
can not be performed.
Consequently, reliable thermal inertia (conductivity) estimates are currently available for a relatively small
number of asteroids \citep[see, e.g.,][]{delbo-etal_2015,harris-drube_2016,2019A&A...625A.139M}. Most often, thermal inertia is derived from analysis of observations in the infrared band. Unless space-borne, such data about very small objects are rare.

The dynamics of bodies up to about 30 kilometers in diameter is known to be influenced 
by thermal effects, caused by solar radiation. Objects are heated up by the Sun, so that they re-radiate away the
energy in the thermal waveband, creating a small thrust and causing a drift 
in semi-major axis, which is perceived over decades: this physical phenomenon is known as 
Yarkovsky effect \citep{1995JGR...100.1585R, rubincam_1998, farinella-etal_1998, bottke-etal_2006, vokrouhlicky-etal_2015}. The Yarkovsky effect perturbs the trajectory,
but it is generated by a thermal force that depends on material properties and the internal structure of
asteroids. Therefore, the Yarkovsky effect is a phenomenon where orbital dynamics is linked to asteroid composition and
physical properties. 

The detection of the Yarkovsky effect has become available as part of orbit determination (i.e., the osculating orbital 
elements of epoch) on a regular basis, in cases where high precision astrometry observations are available. So far this applies uniquely to near-Earth asteroids. 
An alternative approach to estimate the Yarkovsky effect
induced drift in the orbital semi-major axis has been formulated for the first time by \citet{nesvorny-bottke_2004}, where
the authors used backward orbit propagation of the Karin cluster members and monitored convergence of the secular angles. The same method has been later applied to several other young asteroid families. However,
this approach provides less accurate results than the direct orbit-determination mentioned above. Additionally, data about small and fast rotating asteroids, relevant in context of our paper, is not available for the main belt families.

Depending on other available information, measurements of the Yarkovsky effect allow the physical properties of small objects to be constrained. \citet{chesley-etal_2003}  were able to deduce an interval of values for the thermal conductivity of (6489) Golevka, compatible with the existence of thin regolith. Additional examples include \citet{farnocchia-etal_2013}, where the authors were able to constrain the values of the thermal inertia of six NEOs for which the obliquity was known. Similarly, using thermal infrared observations it is possible to estimate the thermal inertia of an asteroid using a suitable thermophysical model \citep[see][and references therein]{delbo-etal_2015}. Combining these measurements with the Yarkovsky drift estimation, it is possible to break the degeneracy between thermal inertia and density, and to directly derive the bulk density of the object \citep{2014Icar..235....5C, farnocchia-etal_2014, mommert-etal_2014,  2014ApJ...786..148M, rozitis-etal_2014,reddy-etal_2016}.

In this paper we developed a statistical method to set constraints on the thermal
conductivity of near-Earth asteroids. Our approach is based on the comparison between the
measured Yarkovsky drift and the theoretically predicted one. Using suitable distributions for the parameters that determine the magnitude of the drift, we perform a Monte Carlo simulation, and produce a probability density function for the thermal conductivity $K$. This probability informs us about the most likely values of $K$, and it can be used to rule out some other values. We applied our model to the asteroid (499998) 2011 PT, a super-fast rotating near-Earth object with a diameter of about 35 meters. Interestingly, we found that the measured semi-major axis drift is large enough such that it can be achieved only for low values of thermal
conductivity, with two most likely values being around 0.0001 and 0.005 W m$^{-1}$ K$^{-1}$, 
suggesting the presence of a thermal insulating layer on the surface of 2011 PT.

\section{Methods}
\label{s:methods}
The semi-major axis drift $da/dt$ due to the Yarkovsky effect for a spherical body placed on a 
circular orbit around the Sun can be expressed through analytical formulas, and it is given by the sum of the 
seasonal and the diurnal effects (see Appendix~\ref{app:analyticYarko}). 
It depends on the orbital parameters, and on the physical and thermal characteristics of the body,
i.e. the semi-major axis
$a$, the diameter $D$, the density $\rho$, the thermal conductivity $K$,
the heat capacity $C$, the obliquity $\gamma$, the rotation period $P$, the absorption coefficient $\alpha$, and
the emissivity $\varepsilon$. Among these parameters, the thermal conductivity $K$ is the most uncertain parameter,
since it strongly depends on the composition of the surface and on the porosity, and it can vary by several 
orders of magnitude. If a measurement $(da/dt)_\text{m}$ of the Yarkovsky effect is available from astrometry \citep[see, e. g.,][]{farnocchia-etal_2013, 2018A&A...617A..61D, greenberg-etal_2020}, solving the model vs observed drift-rate relation
\begin{equation}
    \left(\frac{da}{dt}\right) (a, D, \rho, K, C, \gamma, P, \alpha, \varepsilon) = \bigg(\frac{da}{dt}\bigg)_\text{m},
    \label{eq:yarkoInvertFormula}
\end{equation}
for $K$, allows us to estimate the thermal conductivity, which can be examined statistically using a Monte 
Carlo method. 
To this purpose, in Eq.~\eqref{eq:yarkoInvertFormula} the measured parameters are either fixed to a specific value if their uncertainty is negligible, 
or they can be modeled assuming a Gaussian error distribution. The parameters for which measurements are not
available, or are subject of large uncertainties, can be modeled using a population-based probability 
density function (PDF).

\subsection{Probabilistic model of the parameters of 2011 PT}
Asteroid (499998) 2011 PT is a near-Earth object discovered by the Pan-STARRS 1 survey
telescope on August 1, 2011. With a semi-major axis of $a = 1.3123$~au and an eccentricity of about 
$e = 0.2147$, it falls in the Amor group, meaning that its perihelion distance
$q = a(1-e)$ satisfies $1.017 \text{ au} < q < 1.3 \text{ au}.$
Hereafter we describe the probabilistic model of the orbital and physical parameters of 2011 PT, used for the computation of the semi-major axis induced drift from the theoretical model of the
Yarkovsky effect, i.e. the left hand side of Eq.~\eqref{eq:yarkoInvertFormula}.
%

%

\subsubsection{Absolute magnitude determination}
\label{ss:absMag}
The absolute magnitude $H$ does not play a role in the analytical model of the Yarkovsky effect of 
Eq.~\eqref{eq:yarkoInvertFormula} (see also Appendix~\ref{app:analyticYarko}), however it is related to the size 
of the object and it is needed to model the albedo distribution (see Sec.~\ref{ss:albedo}). 
The JPL Small-Body Database\footnote{\url{https://ssd.jpl.nasa.gov/sbdb.cgi}} (SBDB) reports a value of $H = 23.9$ mag, 
but it is known that orbital catalogs all suffer uncertainties in $H$, including systematic effects 
\citep{pravec-etal_2012}. To overcome this issue and improve the accuracy of the results, we remeasured 
all available Pan-STARRS images and recomputed an absolute magnitude of $H = 24.07 \pm 0.42$ mag
using the phase angle computed from 
the orbital geometry at each observation and the corrections to $V$ band as given by \citet{Denneau2013}. 
We note that the available data does not allow for a more accurate determination of $H$.

\subsubsection{Albedo distribution}\label{ss:albedo}
Although the albedo does not enter directly in the formulation of the Yarkovsky effect as given by
Eq.~\eqref{eq:yarkoInvertFormula}, it is related to both the density and the diameter of the object. 
Indeed, the value $p_V$ of the albedo gives a first indication of the taxonomic 
type, which determines the density, and it provides an estimation of the diameter by the 
conversion formula \citep[see, e.g.,][]{bowell-etal_1989, 2007Icar..190..250P} 
\begin{equation}
   D = \frac{1329 \text{ km}}{\sqrt{p_V}}10^{-H/5},
   \label{eq:mag2dia}
\end{equation}
where $H$ is the absolute magnitude.
A PDF of the albedo of a specific near-Earth object can be constructed 
making use of the NEOs population model by \citet{granvik-etal_2018} and the NEOs albedo distribution
by \citet{morbidelli-etal_2020}.

Given the orbital elements $(a,e,i)$ and the absolute magnitude $H$, the population model by
\citet{granvik-etal_2018} provides the 
probability for a NEO to come from a specific source region, which are namely: the $\nu_6$ secular 
resonance, the 3:1, 5:2 and 2:1 Jupiter mean-motion resonances, the Hungaria region, the Phocaea region, and the
Jupiter Family Comets (JFC). For the sake of notation, we denote this probability with $P_s(a,e,i,H)$, where the
subscript $s$ is used to denote a specific source route. Numerical values corresponding to (499998) 2011 PT are given in Table~\ref{tab:probSR}.

\begin{table}[!ht]
    \caption{The probability to come from each source region according to the model of \citet{granvik-etal_2018}, 
    computed for the orbital elements and absolute magnitude corresponding to the object (499998) 2011 PT.}
    \centering
    \begin{tabular}{cc}
         \hline
         \hline
        Source Region &  $P_s(a,e,i,H)$   \\
         \hline
        $\nu_6$  & 0.67623   \\
        3:1      & 0.08397   \\
        5:2      & 0.0\hphantom{0000}      \\
        Hungaria & 0.23854   \\
        Phocaea  & 0.00013   \\
        2:1      & 0.00113    \\
        JFC      & 0.0\hphantom{0000}      \\
         \hline
    \end{tabular}
    \label{tab:probSR}
\end{table}

In the NEOs albedo distribution by \citet{morbidelli-etal_2020}, objects are binned in three albedo categories:
\begin{itemize}
    \item category 1: $p_V \leq 0.1$;
    \item category 2: $0.1  < p_V  \leq 0.3$;
    \item category 3: $0.3 < p_V$.
\end{itemize}
For each source region, the model provides the fraction of delivered bodies belonging to each albedo category 
$c_1, c_2, c_3$. For the sake of notation, we denote the fractions with $p_s(c_i), i=1,2,3$ where
$s$ indicates again the source region. The exact numerical values are reported in \citet{morbidelli-etal_2020}.
The Hungaria region mostly provides high albedo objects, in agreement with the fact that this part of the 
main belt contains a large number of E-type asteroids \citep{demeo-carry_2013, demeo-carry_2014}. 
The inner belt regions, i.e. the $\nu_6$ secular resonance and the 3:1 Jupiter mean motion resonance, 
mostly provide asteroids in the S-complex. The only exception is the Phocaea region which supplies 
an approximately equal number of dark (category 1) and bright asteroids (categories 2 and 3). A larger fraction
of dark asteroids coming from the Phocaea region compared to the rest of the inner main belt is probably due to a dark asteroid family located therein \citep[][]{2017AJ....153..266N}. Nevertheless, the probability for 2011~PT to originate in this region is very small.
The mid and outer belt regions, i.e. the 5:2 and 2:1 Jupiter mean motion resonances,
account for a higher percentage of C-complex asteroids, according to the fact that primordial carbonaceous 
bodies are more abundant at larger semi-major axis values \citep{demeo-carry_2013, demeo-carry_2014}.

\begin{figure*}[!ht]
   \centering
   \includegraphics[width=1\textwidth]{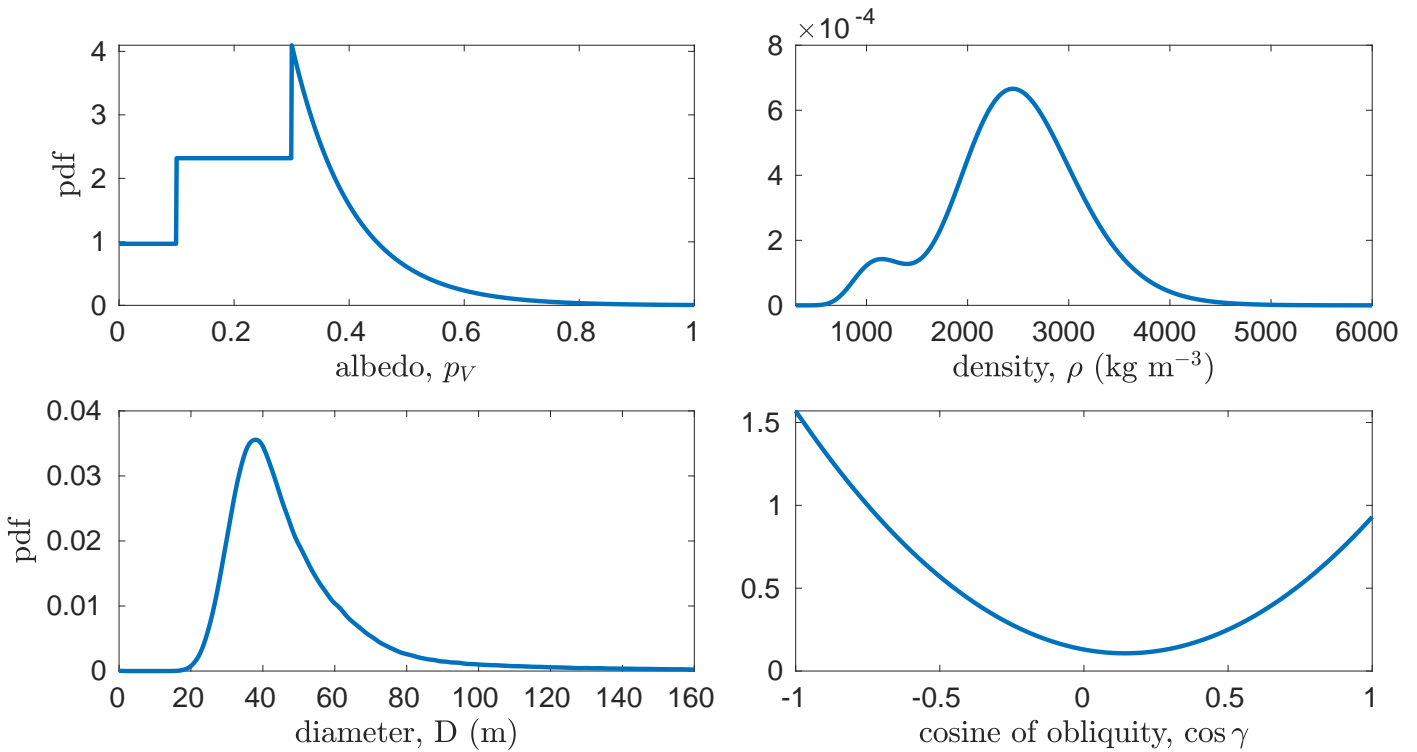}
   \caption{Probability density function for the albedo $p_V$ (top left panel), the density $\rho$ (top right panel), the diameter $D$ (bottom left panel), and the 
   obliquity $\gamma$ (bottom right panel), computed for asteroid (499998) 2011 PT.}
   \label{fig:input_pdf}
\end{figure*}

For each escape route, we define an albedo probability function $p_s(p_V)$ assigning a uniform 
density in the first and second categories, and an exponentially decaying density in the third category, 
as done in \citet{morbidelli-etal_2020}:
\begin{equation}
    p_s(p_V) = 
    \begin{cases}
    \displaystyle \frac{p_s(c_1)}{0.1}, & p_V \leq 0.1,    \\[2ex]
    \displaystyle \frac{p_s(c_2)}{0.2}, & 0.1<p_V\leq 0.3, \\[2ex]
    \displaystyle p_s(c_3)\frac{2.6^{-\frac{p_V-0.3}{0.1}}}{\int_{0.3}^1 2.6^{-\frac{x-0.3}{0.1}}dx}, & p_V > 0.3.
    \end{cases}
    \label{eq:pspv}
\end{equation}
Given the probabilities $P_s(a,e,i,H)$ for the object to come from each source region, 
the albedo PDF $p(p_V)$ for a specific NEO is defined by
\begin{equation}
    p(p_V) = \sum_{s=1}^7 P_s(a,e,i,H)\, p_s(p_V).
    \label{eq:albedo_pdf}
\end{equation}
The resulting albedo PDF for 2011 PT is shown in the top 
left panel of Fig.~\ref{fig:input_pdf}. 


\subsubsection{Density distribution}\label{ss:density}
In order to obtain the PDF of the density of 2011 PT, we first divide asteroids in three main complexes: 
1) the C-complex, composed by dark carbonaceous objects, which includes C-, P-, D-, B-, and T-types,
2) the S-complex, composed by stony objects, which gathers together S-, Q-, V-, A-, and R-types,
and 
3) the X-complex, composed by objects with different compositions, namely the intermediate albedo M-type,
and the high albedo E-type.

For the S-complex we take as a reference the density of S-type, since they are the most common and 
numerous one \citep{demeo-carry_2014, demeo-etal_2015}, which is reported to be
$\rho = 2720 \pm 540 \text{ kg m}^{-3}$ \citep{carry_2012}.

On the other hand, C-type asteroids are the most common in the C-complex, hence we assume their density to model 
this population. \citet{carry_2012} reports a density of 
$\rho = 1330 \pm 580 \text{ kg m}^{-3}$ for the C-type, which however seems to be a bit high if compared to the
precise measurements performed on asteroids (101955) Bennu and (162173) Ryugu. 
(101955) Bennu is a B-type asteroid with a composition similar to the aqueously altered CM 
carbonaceous chondrite, and its density is $1190\pm13$ kg m$^{-3}$ \citep{2019NatAs...3..352S}, while (162173)
Ryugu is a Cb-type asteroid with a composition similar to carbonaceous chondrites, and a density of $1190\pm20$ 
kg m$^{-3}$ \citep{watanabe-etal_2019}. These two precise measurements suggest that the value of the density of C-type asteroids proposed in \citet{carry_2012} may be too large, 
and this deviation might be due to large errors introduced by the method used to produce the estimates. The difference may also be because asteroids tend to have larger density with increasing diameter \citep[see, e.g.,][]{carry_2012, 2014Icar..239...46M}.
For this reason, we assume a lower value for the C-complex, setting the density to
$\rho = 1200 \pm 300 \text{ kg m}^{-3}$.

\begin{figure}[!ht]
   \centering
   \includegraphics[width=0.45\textwidth]{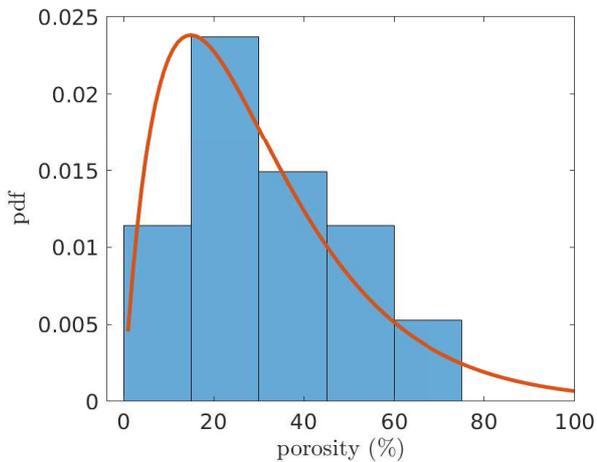}
   \caption{Fit of the asteroid porosity with a gamma distribution, using data from
      \citet{carry_2012}.}
   \label{fig:porosity_fit_Xe}
\end{figure}

As previously said, the X-complex shows a different variety of densities and compositions, however the E-type, 
which is connected to Aubrite meteorites, is the most common type delivered by the Hungaria region 
\citep{2014Icar..239..154C,binzel-etal_2019}, which is one of the main near-Earth objects 
source regions \citep{granvik-etal_2018}. Hence, we assume the density of E-type asteroids as a 
reference for the X-complex.
\citet{carry_2012} reports only one measurement with an accuracy better than 20\% for this asteroid type, and it is equal 
to $2600 \pm 200$ kg m$^{-3}$. However, drawing a statistical picture from one single measurement may not be correct: 
for this reason we established other values. To this purpose, we took into account the grain density of the connected 
Aubrite meteorites, which is reported to be 
$\rho = 3200 \pm 80 \text{ kg m}^{-3}$ \citep{ostrowsky-bryson_2019}, and we corrected it taking into account the porosity. 
The porosity distribution is deduced from the values reported in \citet{carry_2012}: we
took all the asteroids with porosity larger than zero and relative accuracy better than 50\% in the density measurements. 
Then, we fitted these data with a gamma distribution, see Fig.~\ref{fig:porosity_fit_Xe}.
Assuming a Gaussian distribution for the density of Aubrite, we corrected it with the porosity using
the formula 
\begin{equation}
   \rho = \rho_{\text{m}}\,\bigg( 1-\frac{P(\%)}{100}\bigg),
   \label{eq:porosity_correction_Xe}
\end{equation}
from \citet{carry_2012}, where $P(\%)$ is the porosity and $\rho_{\text{m}}$ is the density of the
meteorite. In this manner, we found a density of $\rho = 2350 \pm 520 \text{ kg m}^{-3}$,
which we use for the purpose of this work.


The three taxonomic complexes could be also linked to the three albedo categories defined by Eq.~(3). In this respect, we assume that category 1, which refers to dark asteroids, corresponds to the C-complex, category 2 corresponds to the 
S-complex, since it refers to moderate albedo values, while category 3 is associated to the X-complex, containing higher albedo objects. This can be used to
construct the PDF for the density of 2011 PT.

To generate a distribution of the density, we first generate a random value of the albedo according to the distribution defined by Eq.~\eqref{eq:albedo_pdf}. Then, based on the generated albedo we determine the albedo category and the associated asteroid complex. Finally, we randomly generate a value of the density according to the density distribution of the corresponding asteroid complex, that is assumed to be log-normal \citep[see, e.g.,][]{spoto-etal_2014, tardioli-etal_2017}. 
The PDF for $\rho$ obtained with this procedure, smoothed 
using the kernel density estimation, is reported in the top right panel of Fig.~\ref{fig:input_pdf}. As expected, for 2011 PT higher density solutions ($>2000~\text{kg m}^{-3}$) are much more likely than the low ones ($<1500~\text{kg m}^{-3}$).

\subsubsection{Size distribution} 
The size of an object is estimated using its absolute magnitude
$H$ and its albedo $p_V$ by the conversion formula of Eq.~\eqref{eq:mag2dia}.
Since we determined an absolute magnitude of $H = 24.07 \pm 0.42$ mag,
the diameter is mostly affected by the uncertainty on the
albedo, and it can vary even of an order of magnitude if the range of $p_V$ is large.

%
The distribution of the diameter $D$ is obtained again by conversion of the albedo distribution
of Eq.~\eqref{eq:albedo_pdf}: random values of $p_V$ are generated first, and then converted in diameter using 
Eq.~\eqref{eq:mag2dia}.
The uncertainty in the determination of the absolute magnitude $H$ is taken into account assuming a Gaussian
distribution of the errors.
The PDF for the diameter of 2011 PT obtained with this procedure, 
smoothed using the kernel density estimation, 
is reported in the bottom left panel of Fig.~\ref{fig:input_pdf}. 
The most likely size suggested by this model is about 35 meters.

\subsubsection{Obliquity distribution} 
The obliquity of an asteroid can be determined from photometric
measurements using the lightcurve inversion method \citep{kaasalainen-torppa_2001, 
kaasalainen-etal_2001, kaasalainen-etal_2002, durech-etal_2015}, or from radar observations
\citep{hudson_1994, hudson-etal_2000, magri-etal_2007}.
If measurements are available, it is possible to model this parameter using a Gaussian
distribution, with mean corresponding to the nominal value and standard deviation
corresponding to the error.
On the other hand, if it is unknown, which is the most usual situation,
we have to rely on general properties of the NEOs population.

It is known that in the NEOs population there is an excess of retrograde rotators
\citep{laspina-etal_2004}, caused by the transport mechanism of objects from the main
belt, driven by the Yarkovsky effect \citep{bottke-etal_2002, granvik-etal_2018}.
A first attempt to derive an obliquity distribution for near-Earth asteroids has been made
by \citet{farnocchia-etal_2013}, where the authors used a four-bin probability density
function to match the observed obliquities. A more sophisticated approach has been used in
\citet{tardioli-etal_2017}, where different distributions have been tested. The best fit
solution is obtained for a quadratic distribution of $\cos\gamma$, namely
\begin{equation}
   p(\cos\gamma) = a \cos^2\gamma + b \cos\gamma + c,
   \label{eq:pdf_cosgamma}
\end{equation}
where the parameters are
$a = 1.12, \, b = -0.32, \, c = 0.13$.
The PDF of Eq.~\eqref{eq:pdf_cosgamma} is shown in Fig.~\ref{fig:input_pdf}, bottom right panel.


\subsubsection{Fixed parameters}

\subsubsection*{Semi-major axis}
The uncertainty in the semi-major axis $a$ of 2011 PT determined from astrometry, as reported 
in the JPL SBDB, is about $9.3 \times 10^{-10}$ au: small changes of $a$ within this uncertainty interval produce negligible changes in the predicted Yarkovsky drift, therefore this parameter is kept fixed to the nominal value.


\subsubsection*{Heat capacity}
The heat capacity $C$ depends on the physical characteristics of the 
object, however its value may vary within a factor of several \citep{delbo-etal_2015}.
Typical assumed values for km-size rocky or regolith-covered main belt asteroids are in the range 600-700 J kg$^{-1}$ K$^{-1}$, 
while for iron-rich objects the value is smaller, of the order of 500 J kg$^{-1}$ K$^{-1}$
\citep[see e.g][]{farinella-etal_1998}. However, the heat capacity increases for increasing temperature, and its value may be higher for near-Earth asteroids, supposedly not exceeding $\sim$1200 J kg$^{-1}$ K$^{-1}$.
In \citet{ostrowsky-bryson_2019}, the authors reported the measured heat capacities for different 
type of meteorites. From their data, we can draw the conclusion that $C$ is larger than 500 J kg$^{-1}$ K$^{-1}$ for stony meteorites, while it is smaller than this threshold value if there is a
high component of metallic material. 
However, it is not clear how to produce a distribution from the typical assumed value for
km-size asteroid. Moreover, data on meteorites are not enough to produce a clear statistical
distribution, they might be biased, and it is not clear how to correct their value when we
consider small asteroids.
For all the above reasons, we keep fixed the value of the heat capacity $C$, and produce the 
results for a few different reasonable values. 

\subsubsection*{Emissivity and absorption coefficient} 
\citet{ostrowsky-bryson_2019} reported the measurements of the emissivity
for 61 meteorites, and all the objects but one have an emissivity in between 0.9 and 1,
with an average value of 0.984. Since we do not expect the Yarkovsky effect to vary much
with respect to this parameter, we fix $\varepsilon$ to the mean value computed with
meteorite measurements.
The absorption coefficient is set to $\alpha=1$ in all our calculations.

\subsubsection{Parameters modeled with errors in measurements}

\subsubsection*{Rotation period} 
\citet{kikwaya-etal_2015} determined the rotation period of 2011 PT using photometric observations,
and it results to be $P = 0.17 \pm 0.05 \text{ hr}$, corresponding to about 11 minutes. Although the small uncertainty 
might produce very small variations on the predicted Yarkovsky drift, we take it into account by assuming a
Gaussian distribution of the errors.

\subsubsection*{Yarkovsky drift measurement} 
The Yarkovsky drift has been determined by \citet{greenberg-etal_2020} with a nominal
value and an uncertainty of 
\[
   \bigg(\frac{da}{dt}\bigg)_{\text{m}} = (-88.44 \pm 14.6) \times 10^{-4} \text{ au My}^{-1},
\]
which we used to perform the Monte Carlo simulation. This parameter is also modeled assuming a Gaussian distribution of the errors.
Note that the JPL SBDB provides an automatic solution of the non-gravitational transverse acceleration
$A_2=(-2.237\pm 0.302)\times 10^{-13}$~au d$^{-2}$, corresponding to $(da/dt)=(-86.9\pm 11.7)\times 10^{-4}$ au~My$^{-1}$, 
while \citet{2018A&A...617A..61D} report a value of $(da/dt)=(-91.3\pm 14.44)\times 10^{-4}$ au~My$^{-1}$, 
both nicely consistent with the value given by \citet{greenberg-etal_2020}. 

\subsection{Monte Carlo simulation}
We generate samples
\begin{gather*}
   \{ (D_h, \rho_h) \}_{h=1, \dots, n_{p_V}}, \\  
   \{ \gamma_j \}_{j=1, \dots n_\gamma}, \\
   \{ P_k \}_{k=1, \dots n_P}, \\
   \{ (da/dt)_i \}_{i=1, \dots n_{\text{Yarko}}},
\end{gather*}
for the parameters modeled with a distribution, following the procedure
described in the previous section, where $n_{p_V}, n_\gamma, n_P, n_{\text{Yarko}}$ 
denote the number of points in each corresponding sample. 
Diameter and density are generated in pairs using the same values of albedo, therefore
taking into account their correlation. In this manner we avoid to produce non-physical 
combinations such as small diameter (obtained by converting a high albedo) and low density 
(obtained by converting a low albedo). 
For a given Yarkovsky drift $(da/dt)_i$ and a possible combination of parameters $(D_h, \rho_h, \gamma_j, P_k)$, 
we numerically invert the relation
\begin{equation}
  \left(\frac{da}{dt}\right) (a, D_h, \rho_h, K, C, \gamma_j, P_k, \alpha, \varepsilon) = \bigg(\frac{da}{dt}\bigg)_i,
   \label{eq:inverseProblem}
\end{equation}
computing the values of the thermal conductivity $K$ for which the measured drift is achieved. In this
manner, for each combination of parameters, we obtain 
\begin{equation}
   K_1^{(h,j,k,i)}, \dots, K_n^{(h,j,k,i)},
   \label{eq:thermalCondOut}
\end{equation}
where $n$ is the number of solutions of Eq.~\eqref{eq:inverseProblem}. Collecting all the
solutions together, we obtain a distribution for the possible values of the thermal
conductivity $K$ of the specific object, from which we can compute the corresponding
probability density function using the kernel density estimator.

\section{Results}
\label{s:results}
%

\subsection{Preliminary constraints on the thermal conductivity}
\begin{figure}[!ht]
   \centering
   \includegraphics[width=0.49\textwidth]{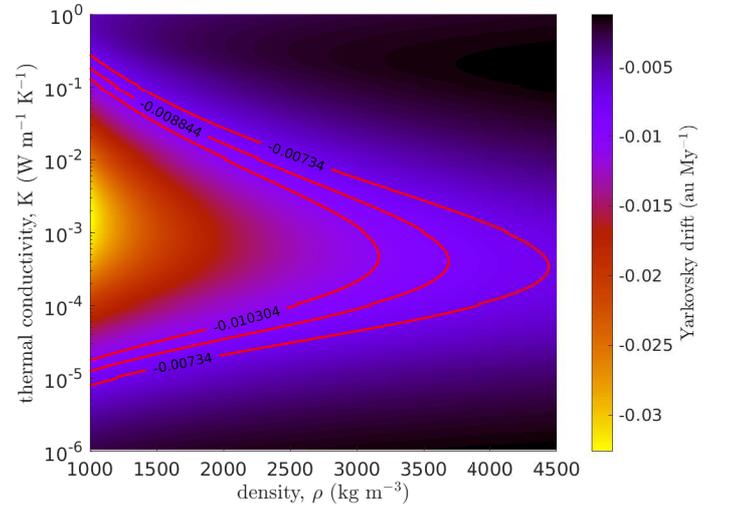}
   \caption{The estimated maximum negative Yarkovsky drift for 2011 PT, obtained for $C = 680$ J kg$^{-1}$ K$^{-1}$, $D = 35$ m, and $P=0.17$ h. 
      The red level curves correspond to the nominal Yarkovsky drift (level curve in the
      between), and the drift obtained taking into account the uncertainties (external level curves).}
   \label{fig:2011PTyarkoMin}
\end{figure}

A first clue that the thermal conductivity of 2011 PT might be small is given by
simply computing the Yarkovsky drift for the nominal values of the parameters. 
For instance, fixing the diameter to $D = 35$ meters, and the heat capacity to
$C = 680$ J kg$^{-1}$ K$^{-1}$, we computed the maximum negative Yarkovsky drift (thus selecting $\gamma=180\degr$) as a function of the
density $\rho$ and the thermal conductivity $K$. The resulting drift values,
together with the three level curves corresponding to the measured nominal value and the
1-$\sigma$ levels, are reported in Fig.~\ref{fig:2011PTyarkoMin}.
Since here we are computing the maximum negative Yarkovsky drift, from the plot of
Fig.~\ref{fig:2011PTyarkoMin} we can conclude that the values of $(\rho, K)$, residing
inside the area determined by the external level curve at 1-$\sigma$ level, are all
compatible with the measurements of the drift.
For all the values of the density we obtained a small value of
thermal conductivity $K$, which is always less than 0.3 W m$^{-1}$ K$^{-1}$, and decreases with
increasing density. For instance, for $\rho = 1500$ kg m$^{-3}$, we have
\[
   0.00001 \text{ W m$^{-1}$ K$^{-1}$} \lesssim K \lesssim 0.074 \text{ W m$^{-1}$ K$^{-1}$},  
\]
while, doubling the density to $\rho = 3000$ kg m$^{-3}$, we obtain
\[
   0.00005 \text{ W m$^{-1}$ K$^{-1}$} \lesssim K \lesssim 0.0055 \text{ W m$^{-1}$ K$^{-1}$}.
\]
These considerations already suggest a low thermal conductivity of the surface,
which has to be analyzed in a more depth taking into account the uncertainties in the 
parameters involved, and by performing the Monte Carlo simulations explained in the previous
section.

We note that the fast rotation of (499998) 2011~PT is a key element of the solution. This is because the maximum drift-rate $da/dt$ is obtained when the diurnal $F$-value is maximum in Eq.~(\ref{eq:yarkoDiurnal}). 
The maximum occurs when $\Theta_{\text{d}}\simeq 1$, namely when the corresponding diurnal thermal parameter is of the order of unity. 
Since $\Theta_{\text{d}}$ depends on the ratio $K/P$ (see Eq.~\ref{eq:thd}), the optimal conductivity $K$
scales as $K\propto P$.
Therefore, the whole pattern seen in Fig.~\ref{fig:2011PTyarkoMin} would slide toward larger conductivity values in the ordinate for more slowly rotating bodies. In the case of the very fast-rotating asteroid 2011 PT, 
$K$ is constrained to values smaller than $\sim$0.1 W m$^{-1}$ K$^{-1}$.

\subsection{Results of the Monte Carlo simulations}
\begin{figure*}
    \centering
    \includegraphics[width=\textwidth]{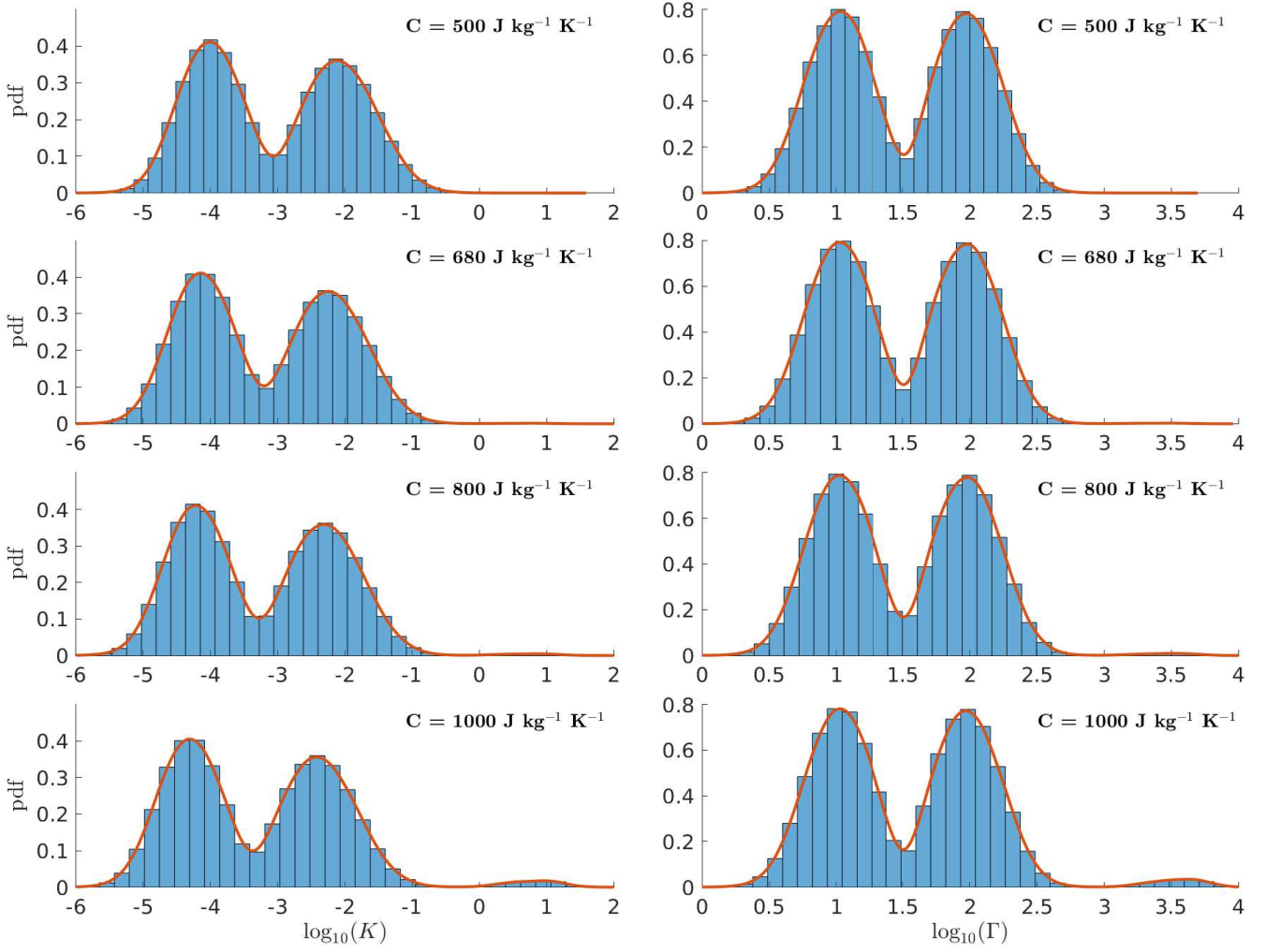}
    \caption{Derived thermal parameters of 2011 PT obtained from the Monte Carlo simulations, for different values of the heat capacity $C$. The plots on the 
    first column show the distribution of the thermal conductivity $K$, while the plots on the second column show the 
    distribution of the thermal inertia $\Gamma$. The thermal inertia is obtained through $\Gamma = \sqrt{\rho K C}$, where $\rho$ 
    is the corresponding value of the density used to invert Eq.~\eqref{eq:inverseProblem}.}
    \label{fig:Kdistribs}
\end{figure*}

To better quantify the low-thermal conductivity of the surface of (499998) 2011 PT, we
performed a Monte Carlo simulation with the method described in Sec.~\ref{s:methods}. 
We generated samples $\{ (D_h, \rho_h) \}, \, \{ \gamma_j \}, \, \{ P_k \}, \, \{ da/dt \}_i $ of 50000 points for diameter and density, and 10000 points for obliquity, rotation period, and Yarkovsky drift, respectively.
To perform the Monte Carlo experiment and generate the results, 
we randomly chose 1 million combinations of the parameters.
The output distribution for the thermal conductivity $K$ was then fitted using the
kernel density estimation, so that to obtain a smooth probability density function. 
We ran different simulations for some values of the heat capacity $C$, namely
\[
   C = 500, \, 680, \, 800, \, 1000, \, 1200 \text{ J kg$^{-1}$ K$^{-1}$}.
\]
Using these values we explore the expected range of $C$ for near-Earth objects, from which we can extrapolate an eventual trend.
The results of the Monte Carlo simulations are shown in Fig.~\ref{fig:Kdistribs}, together with the corresponding 
distribution obtained for the thermal inertia 
\begin{equation}
    \Gamma = \sqrt{\rho K C}.
    \label{eq:thermalInertia}
\end{equation}
From the plots we can notice that we always obtained a bimodal distribution, with a first peak in
the interval
\[
   0.00001 \text{ W m$^{-1}$ K$^{-1}$} < K < 0.001 \text{ W m$^{-1}$ K$^{-1}$}, 
\]
and the second one in the interval
\[
   0.001 \text{ W m$^{-1}$ K$^{-1}$} < K < 0.1 \text{ W m$^{-1}$ K$^{-1}$}.
\]
The distributions of the thermal inertia $\Gamma$ are also bimodal, with a first peak in the interval
\[
   5 \text{ J m$^{-2}$ K$^{-1}$ s$^{-1/2}$} < \Gamma < 20 \text{ J m$^{-2}$ K$^{-1}$ s$^{-1/2}$}, 
\]
and the second one in the interval
\[
   75 \text{ J m$^{-2}$ K$^{-1}$ s$^{-1/2}$} < \Gamma < 150 \text{ J m$^{-2}$ K$^{-1}$ s$^{-1/2}$}. 
\]
Moreover, as the heat capacity $C$ increases, a third peak started to grow at large 
values of thermal conductivity, maintaining however a small probability.
The bimodal distribution is explained by the fact that, for each set of
parameters $\{ (D_h, \rho_h) \}, \, \{ \gamma_j \}, \, \{ P_k \}, \, \{ (da/dt)_i \}$, there are always two 
low thermal conductivity solutions of Eq.~\eqref{eq:yarkoInvertFormula}, as can be easily noticed from 
Fig.~\ref{fig:2011PTyarkoMin}. 
Furthermore, since the input distributions all have only one most likely value, most of the combinations of 
input parameters selected during the Monte Carlo simulations are gathered nearby a most likely one,
and this produces two well distinct peaks of low $K$. 
On the other hand, the third peak with small probability is caused by other high thermal 
conductivity solutions of Eq.~\eqref{eq:yarkoInvertFormula} appearing at high heat capacity, 
for large diameter (around $\sim$60 meters) and small density (around $\sim$1100 kg m$^{-3}$). 
These combinations of density and diameter are generated with small probability, 
since they correspond to the low albedo case.

\begin{figure}[htbp]
   \centering
   \includegraphics[width=0.45\textwidth]{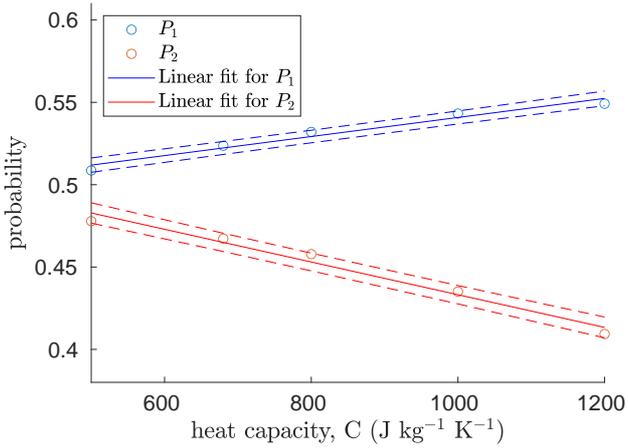}
   \caption{Linear fits of the probabilities $P_1$ and $P_2$ with respect to
   the heat capacity $C$. The dashed lines represent the 99\% confidence interval region.}
   \label{fig:linFit}
\end{figure}

To give quantitative boundaries for the most likely values of thermal conductivity, we computed 
the probabilities 
\begin{equation}
   \begin{split}
      P_1 & = P(0.00001 < K < 0.001), \\
      P_2 & = P(0.001 < K < 0.1),
   \end{split}
   \label{eq:peakProbabilities}
\end{equation}
for the different simulations, as well as their sum and the two local maxima $K_1, K_2$. 
The results are reported in
Table~\ref{tab:outProbabilities}. 
\begin{table}[t]
   \caption{The probabilities $P_1, \, P_2$ of Eq.~\eqref{eq:peakProbabilities}, the
   corresponding sum, and the locations of the two local maxima $K_1, \, K_2$, for the
   different Monte Carlo simulations. } 
   \centering
   \begin{tabular}{cccccc}
         \hline
         \hline
       $C$ & $P_1$ & $P_2$ & $P_1+P_2$ &
      $K_1$&
      $K_2$\\
         \hline
 500 & 0.50860&  0.47788&   0.98649 & $1\times10^{-4}$ & 0.0076 \\
 680 & 0.52367&  0.46725&   0.99092 & $7\times10^{-5}$ & 0.0057 \\
 800 & 0.53200&  0.45785&   0.98985 & $6\times10^{-5}$ & 0.0048 \\
 1000& 0.54317&  0.43511&   0.97828 & $5\times10^{-5}$ & 0.0037 \\
 1200& 0.54916&  0.40935&   0.95852 & $4\times10^{-5}$ & 0.0032 \\
         \hline
   \end{tabular}
         \tablefoot{The heat capacity in the first column is expressed in J kg$^{-1}$ K$^{-1}$, and the
   thermal conductivity in the fifth and sixth columns are expressed in W m$^{-1}$ K$^{-1}$.}
   \label{tab:outProbabilities}
\end{table}
From these values, and from the plots of Fig.~\ref{fig:Kdistribs}, we can conclude that 
the probability for the thermal conductivity of being larger
than 0.1 W m$^{-1}$ K$^{-1}$ was always almost zero, 
with slightly larger values for larger values of heat
capacity. The highest peak $K_1$ lies always in the first interval and it is around
0.0001 W m$^{-1}$ K$^{-1}$, while the lowest one $K_2$ belongs to the second, being of the order of
$\sim$0.005 W m$^{-1}$ K$^{-1}$. Moreover, we can notice that they both move to lower values of thermal 
conductivity as the heat capacity $C$ increases. The probability $P_1$ results to be always 
greater than the probability $P_2$, but only with a few percent in difference. 
The values of $P_1$ and $P_2$, as function of the heat capacity $C$, can be fitted linearly and the results
are show in Fig.~\ref{fig:linFit}, together with the 99\% confidence interval region of the fit. 
As the heat capacity increases, the probability $P_1$ increases, while $P_2$ decreases, with a more pronounced slope than $P_1$,
which results in a decreasing trend in their sum (see Table~\ref{tab:outProbabilities}). This fact was however expected,
due to the presence of the third peak growing at large thermal conductivity.

\section{Discussion}

\subsection{Robustness and limitations of the model}

\subsubsection{Uncertainty due to the orbital eccentricity}
The semi-major axis drift due to the Yarkovsky effect described analytically 
in Appendix~\ref{app:analyticYarko}
refers to an object revolving around the Sun on a circular orbit. However, the Yarkovsky effect may be increased on elliptic orbits \citep{spitale-greenberg_2001}, and it may cause an object to reach the measured migration rate even for larger values of thermal conductivity. While significant changes are expected for orbits with high eccentricity, it is interesting to test the role of moderately small orbital eccentricity in the case of (499998) 2011~PT.

The osculating semi-major axis drift due to the Yarkovsky effect for an asteroid placed on an eccentric orbit is given by
\begin{equation}
    \frac{da}{dt} = \frac{2}{n^2 a} \, \vec{f}_Y \cdot \vec{v},
    \label{eq:yarkovskyWithEccentricity}
\end{equation}
where $n$ is the mean motion, $\vec{v}$ is the heliocentric orbital velocity and $\vec{f}_Y$ is the instantaneous 
value of the Yarkovsky acceleration. 
The acceleration $\vec{f}_Y$ can be computed analytically, for instance, using the formulation given in \citet{vokrouhlicky-etal_2017}, including 
both the diurnal and the seasonal effects (again with the simplifying assumption of linearization of the boundary conditions). To compute the total semi-major axis drift we average the values of the
instantaneous drift rate of Eq.~\eqref{eq:yarkovskyWithEccentricity} over an orbital period numerically (rather than analytically as before for the circular orbit). 

To test the effects of the orbital eccentricity on our results, we compared the mean drift rates obtained for the circular orbit and the eccentric orbit, performing again a Monte Carlo simulation. 
Obviously, we chose the eccentricity $\sim$0.215 appropriate for (499998) 2011~PT. 
We used the same distributions of Sec.~\ref{s:methods} for the diameter, the density, the obliquity, and the rotation period. 
Moreover, we produced a sample of the thermal conductivity using the distributions obtained in Sec.~\ref{s:results}. 
Using a million combinations of the parameters, we evaluated the two different drift rates and computed the relative difference. 

We applied this procedure to (499998) 2011 PT, using a fixed heat capacity of $C = 680$ J kg$^{-1}$ K$^{-1}$. 
The distribution of the percent absolute differences had two peaks, one around 1\% and the other one at 10\%, and almost all the combinations gave a relative difference smaller than about 10\%.
Moreover, the peak at 1\% was found to be corresponding to the lower peak in thermal conductivity, 
while the peak at about 10\% was associated to the second peak in $K$, 
meaning that a higher conductivity produces larger changes
in the Yarkovsky drift when the eccentricity of the orbit is taken into account.
The contribution of the eccentricity to the total semi-major axis drift is therefore limited, and the circular 
orbit approximation should not produce misleading results for the thermal conductivity. 

\subsubsection{Further issues of the Yarkovsky effect modelling}
We also briefly discuss further issues related
to the Yarkovsky effect model outlined in Appendix~\ref{app:analyticYarko}
and later used in Sec.~\ref{s:results}. The Yarkovsky effect
in real-world objects is more complicated in several aspects not yet explored above.
These aspects include (i) non-linearity of the boundary conditions
for heat conduction at the surface, (ii) non-isotropy of the
thermal emission by the surface facets ("thermal beaming"),
(iii) non-sphericity of the shape. While we did not test
any of them in detail, we note that luckily some of them may
affect the result in opposite ways, thus potentially compensating each other
(at least partially), while others may play a minimal role for (499998) 2011~PT.

As an example we note that the non-linearity effects decrease
the theoretically predicted semi-major axis drift rate
$da/dt$ in the relevant range of thermal conductivity values
\citep[see already][]{1995JGR...100.1585R, capek2007}.
The reduction factor ranges between $0.7$ and $0.9$. On the
other hand, thermal beaming effects always increase the
semi-major axis drift rate by a factor ranging between $1.1$
to $1.5$ \citep[see, e.g.,][]{2012MNRAS.423..367R}. The two effects
thus have a tendency to compensate each other, though obviously
in particular cases some residual factor may remain. At large
obliquity, which we consider the most probable for 2011~PT,
the non-sphericity decreases the predicted semi-major axis drift
rate \citep[see, e.g.,][]{1998A&A...338..353V}. In fact, at this
configuration the body presents smaller cross-section
to the sunlight, and thus processes less thermal energy if
compared to a spherical body. The reduction factor may be
similar to that characterising the non-linearity effects. While
we do not have a shape model of 2011~PT, the low amplitude of the
available photometric lightcurve in \citet{kikwaya-etal_2015}
suggests a near-spherical shape; or, better,
taken the opposite way there is no suggestion of largely
non-spherical shape of 2011~PT. Thus it is possible that 
non-sphericity effects in the evaluation of the Yarkovsky drift rate are only small \citep[modifying the predicted semimajor axis drift not more than a factor $\sim (0.8-0.9)$; see Fig.~3 in][]{1998A&A...338..353V}.

Another set of real-world details that were not included in our simple analytical model of the Yarkovsky effect has to do with physical parameters. In particular, they were assumed uniform and constant. In reality, things may be more complicated. For instance the thermal conductivity, and to a lesser extend also the heat capacity, may be temperature dependent. This is especially true for powdered mixtures of minerals (such as the proposed regolith-like surface), because energy transfer by conduction at the contact of the grains is efficiently accompanied by radiative transfer through cavities. Luckily, the orbital eccentricity of (499998) 2011~PT is small, such that the thermal conditions at perihelion and aphelion do not change dramatically. For low-enough conductivity they may not exceed the temperature variations of the daily cycle. For instance the equilibrium subsolar temperature at pericenter and apocenter are $\sim$320~K and $\sim$390~K, respectively. Data in \citet{gundlach-blum_2013} indicate that the conductivity would typically change in this temperature range by a factor $\sim$1.5-1.7 only. So our assumed constant values should represent a median within such a range of variation. While the true effects of varying conductivity are not modeled here, we expect their effect remains small. 

Yet another issue is the assumption of homogeneity of the physical parameters. Consider, for instance, the density $\rho$ discussed in Sec.~\ref{ss:density}. The way we introduced it corresponds to the volume-average (bulk) density. However,
advocating for the existence of the regolith layer, we should
ideally adopt a more complex model of the body: a sphere with a core having the bulk density and a lower-density layer at the surface. For the fast rotation, the dominant diurnal effect remains constrained to the skin processes. Therefore the density in the right hand side of Eqs.~\eqref{eq:penetrationDepth}, \eqref{eq:ths} and \eqref{eq:thd} should be the surface density, rather than the bulk density. Because the surface density is assumed to be smaller (or at maximum equal) to the bulk density, the same thermal lag (and thus also dynamical effect) would need larger conductivity. The scaling is not a priori clear. In the limit of the regolith depth much larger than the penetration depths of diurnal and seasonal thermal wave $l_d$ and $l_s$,
$K$ recalibrates proportionally to the ratio of the bulk density over the surface density (this is because the thermal parameter in Eqs.~\ref{eq:ths} and \ref{eq:thd} contains a $K\rho$ factor). Since the plausible density ratio may peak at $\sim$2, the required conductivity values would be skewed to somewhat larger values. For instance the nicely symmetric solution pattern seen in Fig.~\ref{fig:2011PTyarkoMin} would shift to larger conductivity values for densities above $\sim$2000 kg~m$^{-3}$. However, even this approach may be too simple when
$l_d$ and, especially $l_s$, become comparable or larger than the regolith depth. In this case a full-fledge two-layer model would be needed. \cite{1999A&A...350.1079V} developed this approach for the seasonal component, however a complete formulation for the diurnal component is not presently available. Therefore, at this moment we can only conclude with noting that the fully consistent thermal model with the low-density regolith layer included may result in a moderate increase of the surface conductivity value.

Despite the modeling issues discussed above, it is worth mentioning that \citet{2014Icar..235....5C} made a prediction of the bulk density of asteroid (101955) Bennu by combining a model for the Yarkovsky effect, based on known physical properties, and the measurements of the semi-major axis drift obtained from orbit determination. The more accurate bulk density value obtained during the OSIRIS-REx mission \citep{2019NatAs...3..352S} revealed that the estimate by \citet{2014Icar..235....5C} was indeed correct. This successful prediction could suggest that the model is accurate enough for the analysis performed on 2011 PT.

\subsection{The low thermal conductivity of (499998) 2011 PT}
The results obtained with the Monte Carlo simulations suggest that the Yarkovsky drift determined for (499998) 2011~PT can be achieved preferentially for small values of thermal conductivity $K$, with two most likely values corresponding to the
local maxima: the first one around 0.0001 W m$^{-1}$ K$^{-1}$ and the second one around 0.005 W m$^{-1}$ K$^{-1}$ (see Table~\ref{tab:outProbabilities}). 
This conclusion can be explained in plain words by the combination of the fast rotation and the relatively large Yarkovsky drift. 
Indeed, if the thermal conductivity were large, the fast rotation would largely spread the thermal gradient on the surface 
of the asteroid, and a large semi-major axis drift would be impossible to achieve. On the other hand, a low thermal 
conductivity would be able to keep a fair difference in temperature between the day-side and the night-side of the asteroid, 
that would be able to produce larger semi-major axis drifts.
The results obtained here indicate that values of thermal conductivity larger than 0.1 W m$^{-1}$ K$^{-1}$ are very unlikely, and the admitted values are instead fully compatible with the presence of a thermal insulating layer on the surface. This suggests that the surface is covered by a regolith-like material or very thin dust.
As a comparison, the values of the thermal conductivity of lunar dusty regolith, obtained during the \textit{Apollo} missions, are
between 0.0012 W m$^{-1}$ K$^{-1}$ and 0.0035 W m$^{-1}$ K$^{-1}$ \citep{cremers-birkebak_1971, cremers-hsia_1974}, which are comparable with the values
of the second peaks reported in Table~\ref{tab:outProbabilities}. Assuming that thermal properties of the surface give indication
about the grain size, \citet{gundlach-blum_2013} found that the radius of lunar regolith is estimated to be about 50 $\mu$m, which is compatible with the size distribution of sample lunar soil \citep{bevan-etal_1991}. According to this, the value of thermal conductivity of the second peak suggests the presence of dusty regolith grains with a diameter of several tens of $\mu$m on the surface of (499998) 2011 PT.

At the same time, the thermal conductivity generally decreases with reducing particle size, and thus the smaller values obtained for the first peak might suggest an even smaller size of the grains. However, the dust layer in practice consists
of particles of different sizes, but the thermal properties of particle size mixtures have not yet been well-studied. Recently, \citet{2020JGRE..12506100R} developed a 3D model to study the thermal conductivity of regolith, by simulating heat flow in randomly packed spheres. These authors found that results predicted by simpler theoretical models are not reliable when the particles themselves are made of a material that itself has relatively low thermal conductivity, that could result in significant underestimation of particle sizes on asteroid surfaces.



It is worth noting that typical values for the thermal conductivity of bare rock are
around 2-5 W m$^{-1}$ K$^{-1}$, however low values of the order of 0.1 W m$^{-1}$ K$^{-1}$
might be obtained also for a very large porosity of the material, since measurements performed
on meteorites show that $K$ decreases as the porosity increases \citep{opeil-etal_2010,opeil-etal_2012, ostrowsky-bryson_2019}. 
\citet{kikwaya-etal_2015}, using photometry data, suggested a classification as a X-type asteroid for (499998) 2011 PT, which would imply a stony composition.
Moreover, the high spin rate rules out
the possibility that this object is actually a pure rubble-pile \citep{pravec-harris_2000},
since it is lower than the cohesionless spin-barrier of 2.2 hr, hence the large porosity hypothesis does not seem very likely. 
On the other hand, it is not fully clear what minimum level of cohesion forces are able to withstand moderate stresses existing on 2011~PT, what the associated level of maximum porosity is and what it implies for thermal inertia of the surface. These issues deserve further analysis in the future.

We also note that the third peak at high thermal conductivity (see Fig.~\ref{fig:Kdistribs}, bottom panels), arising at large heat capacity $C$, would either suggest a bare rock composition or the presence of irony material on the surface. 
On the other hand, the presence of iron on the surface would cause a decrease of the heat capacity, as shown by measurements performed on meteorites collected in \citet{ostrowsky-bryson_2019}, which is in contradiction with the fact that this peak grows at large $C$, leaving therefore only the bare rock composition compatible with the third peak solution. However, the occurrence of a high thermal conductivity, associated to the third peak, is very unlikely due to the low probability.
%
According to the above facts, the hypothesis of the presence of dusty regolith-like material seems the most probable
one.

\citet{harris-drube_2016} analyzed the published values of thermal inertia $\Gamma$ estimated through thermophysical models \citep{delbo-etal_2015}, and suggested a possible correlation between the thermal inertia and the rotation period of NEOs, such that $\Gamma$ is decreasing for increasing spin rates. On the other hand, \citet{2019A&A...625A.139M} found no evidence of this trend at very slow rotation rates. Assuming that the trend of \citet{harris-drube_2016} is valid also for shorter rotation periods, our result would support this hypothesis, though
we underline that data used by \citet{harris-drube_2016} and \citet{2019A&A...625A.139M} refer to objects with spin period larger than 2 hr, corresponding to the spin barrier of rubble-piles. Moreover, it is difficult to draw any reliable conclusion about the trend for super-fast rotators, at the moment, since there are evidences that such rotators could have also a high thermal inertia, as the reported estimate for asteroid (54509) YORP indicates \citep{delbo-etal_2015}. 


\subsection{Regolith production and retention on small bodies}
The low thermal conductivity we found for the surface of (499998) 2011 PT could be explained with the presence of regolith. We discuss here the usual mechanisms that are capable to produce such material, and give some general hypotheses of how it may be retained on the surface of a fast rotator.

It is generally thought that regolith can be produced in different ways. One of the main hypotheses is by micrometeoroid impacts, which produce small grains on the surface by means of fallback ejecta and by breaking up of boulders \citep{hoerz-etal_1975, horz-cintala_1997}. However, the velocity of crater ejecta is much larger than the escape velocity of small asteroids \citep{housen-holsapple_2011}, making this process not very feasible in this context.  
A second possibility is by thermal fatigue, proposed by \citet{delbo-etal_2014}. The temperature on the surface of an asteroid follows a diurnal cycle, and the stress resulting from sudden changes of temperature can produce damages to the surface material. This mechanism of rock weathering and cracking without ejection could be a possible explanation for the presence of regolith on small bodies. Moreover, thermal fatigue is a process which is independent from the size of the asteroid, as opposite from the micrometeoroid impacts process. The cracking induced by the constant day-night cycle has been recently observed on asteroid (101955) Bennu, with in-situ images taken during the OSIRIS-REx mission \citep{molaro-etal_2020}.
Another hypothesis is that the asteroid itself is active. An active asteroid holds some mechanism that causes mass shedding and ejection of particles from the surface, such as volatile releasing, rotational disruption, ice sublimation, and electrostatic lofting \citep{2015aste.book..221J}. Particles may be ejected from the surface with low-enough velocity and consequently re-impact on the asteroid again, contributing to the cracking of superficial material due to micro-impacts. This process of particle ejection and fallback has been observed during the OSIRIS-REx mission \citep{lauretta-etal_2019b, mcmahon-etal_2020}, but it is not clear if such activity is limited to primitive bodies of spectral types within the C-complex (which would not be applicable for 2011~PT).

To explain the presence of regolith on small and fast rotators, \cite{sanchez-scheeres_2020} developed a model to study under which rotational conditions the surface material is retained, partially lost, or completely lost. The authors found that the covering material is preferentially lost across some regions of the body, and the area grows as the spin rate increases. Nevertheless, regions at high latitudes are able to keep regolith at arbitrarily high spins. This poses a clue that even small and very fast rotators can retain regolith-like grains and dust on their surface. However, it remains to
be seen what happens to these polar seed-regions when the Yarkovsky-O'Keefe-Radzievskii-Paddack (YORP) effect \citep{vokrouhlicky-etal_2015} causes changes in the rotation rate, and potentially also orientation of the spin axis in space. A robust retention process of the formed grains on the surface may be needed.

\begin{figure}[ht]
    \centering
    \includegraphics[width=0.48\textwidth]{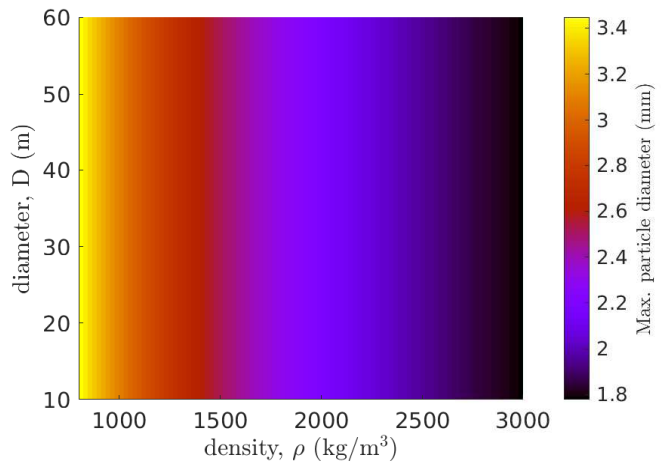}
    \caption{The maximum grain size, as a function of the density $\rho$ and the diameter $D$. Assuming a bond number equal to ten in Eq.~\eqref{eq:choesiveForce}, the maximum grain size allowed refers to the value obtained for the maximum ambient gravity $g_A$ over the asteroid surface.}
    \label{fig:maxGrainSize}
\end{figure}

Such a retention phenomenon may be provided by some kind of cohesive forces, such as van der Waals forces, produced by the direct interaction of solar plasma with surface material, creating a superficial 
electric field. Cohesion may be able to hold fine grains together and stick them to the surface, and they have been
demonstrated  to be very high in lunar regolith \citep{mitchell-etal_1974}. The strength of the cohesive force is represented by 
the bond number $B$, which is given by
\begin{equation}
    B = 10^{-5} g_A^{-1} d^{-2},
    \label{eq:choesiveForce}
\end{equation}
where $g_A$ is the ambient gravity and $d$ is the grain diameter \citep{scheeres-etal_2010, rozitis-etal_2014}. The stability 
of surface material needs a bond number larger than ten \citep{rozitis-etal_2014}, hence we can set upper constraints
on the diameter of grains. Assuming the nominal rotation period and an obliquity of $170\degr$, we computed the ambient 
gravity $g_A$ as a function of the diameter $D$ of the asteroid and the density $\rho$, using a spherical model for the shape. 
Then, we computed the upper limit for the grain diameter $d$ setting a bond number equal to ten in Eq.~\eqref{eq:choesiveForce}, obtaining a function of $R$ and $\rho$. The result is shown in Fig.~\ref{fig:maxGrainSize}.

The maximum grain diameter obtained is between 1 and 4 millimeters. Moreover these bounds are essentially independent of the obliquity $\gamma$ between $90\degr$ and $180\degr$. These estimates give upper bounds for the size of particles that can be held on the surface by means of cohesive forces, and they are well above the $\sim$10-100 $\mu$m size of the lunar regolith expected from the second peak of thermal conductivity. 
Therefore, the hypothesis of the existence of cohesive forces is a reasonable explanation, which could further be explored using better shape and thermophysical models, in case that additional appropriate observational data will be acquired.

\section{Summary and conclusions}
In this paper we developed a statistical method to constrain the thermal conductivity
$K$ of (499998) 2011 PT, a near-Earth object of about 35 meters in diameter which rotates very fast, with a period of 11 minutes. The method is based on the comparison between the
measured Yarkovsky drift and the theoretically predicted one, which depends on the orbital and 
physical parameters of the asteroids.
%
%
%
Modeling distributions for the input parameters, we used a Monte Carlo simulation
to produce a distribution of values of the thermal conductivity, which are compatible 
with the measured Yarkovsky drift.
When the sample is obtained, a probability density function for $K$ was computed, 
and it was used to deduce the most likely values. The method developed here is very flexible with respect to the known and unknown input parameters, and it can be easily applied to asteroids with different size and rotation period. 
It may be also customized on a case by case basis, in order to obtain the best possible solution in each individual case.

The obtained probability density functions for the thermal conductivity of 2011 PT always have two local maxima, one around 0.0001 W m$^{-1}$ K$^{-1}$, and the another one around 0.005 W m$^{-1}$ K$^{-1}$. These two values are the most likely ones, with the first one having a slightly higher probability. With the high probability of at least 95 per cent, we constrained the thermal conductivity to be in the range 0.00001 W m$^{-1}$ K$^{-1}$ < $K$ < 0.1 W m$^{-1}$ K$^{-1}$. 
This is the first case that the low thermal conductivity solution for a small and super-fast rotating asteroid has been obtained with such high probability. The low-thermal conductivity can be interpreted as the presence of regolith-like 
material or thin dust on the surface of (499998) 2011 PT, opening many related scientific questions, and in particular how these super-fast rotators are able to hold such material on their surface.

%
%

To better explore this possibility, however, more observations and characterizations of small asteroids are needed.
A more precise answer can be obtained by planning space missions to one or more such NEOs, 
developing eventually a technique to sample some surface material of fast rotators, which can be brought
back to Earth for an accurate analysis. 
In this respect, the proposed extension of the JAXA Hayabusa~2 mission to visit the asteroid\footnote{
\url{http://www.hayabusa2.jaxa.jp/enjoy/material/press/Hayabusa2_Press_20200915_ver9_en2.pdf}
} 1998 KY26, which is about 30 meters in diameter and rotates with a
period of about 10 minutes, may provide very interesting insights on the structure and the composition of very small and fast rotating objects.

   \begin{acknowledgements}
  We thank the anonymous referee for the suggestions that helped us to improve the manuscript.
   M. F. and B. N. have been supported by the
MSCA-ITN Stardust-R, Grant Agreement n. 813644 under the European Union H2020 research and innovation
program. The work of D. V. was partially funded by the Czech Science Foundation (grant 18-06083S).

This research has made use of data and/or services provided by the International Astronomical Union's Minor Planet Center. 

Pan-STARRS is supported by the National Aeronautics and Space Administration under Grant No. 80NSSC18K0971 issued through the SSO Near Earth Object Observations Program.

   \end{acknowledgements}

    \bibliographystyle{aa}
    \bibliography{stardustBib.bib}{} 

    \appendix

\section{Analytical formulation of the Yarkovsky effect}
\label{app:analyticYarko}
We recall here the analytical formulation of the Yarkovsky effect given in, for instance, \citet{1998A&A...338..353V, vokrouhlicky_1999}.
We assume to have a spherical body moving on a circular orbit around the Sun, that rotates along a fixed axis.
By linearizing the boundary condition of the heat diffusion problem, the drift in semi-major axis $a$ due to the Yarkovsky effect
is given by two distinct components: the seasonal effect
\begin{equation}
    \bigg( \frac{da}{dt} \bigg)_{\text{s}} =\hphantom{-} \frac{4 \alpha}{9}
   \frac{\Phi}{\omega_{\text{rev}}} F(R'_{\text{s}}, \Theta_{\text{s}})\sin^2\gamma, 
   \label{eq:yarkoSeasonal}
\end{equation}
and the diurnal effect
\begin{equation}
    \bigg( \frac{da}{dt} \bigg)_{\text{d}} = -\frac{8 \alpha}{9}
   \frac{\Phi}{\omega_{\text{rev}}} F(R'_{\text{d}}, \Theta_{\text{d}})\cos\gamma. 
      \label{eq:yarkoDiurnal}
\end{equation}
In the above equations, $\alpha$ is the absorption coefficient of the surface, $\Phi$ is
the radiation pressure coefficient \citep[e.g.,][]{vokrouhlicky-etal_2015}, $\omega_{\text{rev}}$ is the orbital frequency (mean motion), and $\gamma$ is the spin axis obliquity. 
$R'_{\text{s}}$ and $R'_{\text{d}}$ are the scaled (nondimensional) values of the radius $R$, defined as
\begin{equation}
  R'_{\text{s}}= \frac{R}{l_{\text{s}}}, \quad R'_{\text{d}} = \frac{R}{l_{\text{d}}},
   \label{eq:rescaledRadius}
\end{equation}
where $l_{\text{s}}, \, l_{\text{d}}$ are the penetration depths of the seasonal and diurnal thermal wave given by 
\begin{equation}
   l_{\text{s}} = \sqrt{\frac{K}{\rho C\omega_{\text{rev}}}}, \quad
   l_{\text{d}} = \sqrt{\frac{K}{\rho C\omega_{\text{rot}}}}.
   \label{eq:penetrationDepth}
\end{equation}
Note that $l_{\text{s}}$ and $l_{\text{d}}$ depend on the thermal conductivity $K$, the heat capacity $C$, and the density $\rho$ of the asteroid. Additionally, the two length scales depend on the respective frequencies: (i) the spin frequency $\omega_{\text{rot}}$ in the case of the diurnal effect, and (ii) the orbital frequency $\omega_{\text{rev}}$ in the case of the seasonal effect.
The thermal parameters $\Theta_{\text{s}}, \Theta_{\text{d}}$ also depend on the physical and thermal characteristics
of the object, and they are defined as 
\begin{gather}
   \Theta_{\text{s}} = \frac{\sqrt{\rho K C \omega_{\text{rev}}}}{\varepsilon \sigma T_\star^3}, 
    \label{eq:ths} \\
   \Theta_{\text{d}} = \frac{\sqrt{\rho K C \omega_{\text{rot}}}}{\varepsilon \sigma T_\star^3}, 
    \label{eq:thd}
\end{gather}
where $\sigma$ is the Stefan-Boltzmann constant, $\varepsilon$ is the emissivity and
$T_\star$ is the subsolar temperature, defined by $\varepsilon\sigma T_\star^4 = \alpha
\mathcal{E}_\star$, with $\mathcal{E}_\star$ being the solar radiation flux at the
distance of the body. The function $F$ in Eqs.~(\ref{eq:yarkoSeasonal}) and (\ref{eq:yarkoDiurnal}) depends on both the corresponding scaled radius and the thermal
parameter, and it determines the total magnitude of the Yarkovsky drift.
The definition of $F$ is given by
\begin{equation}
    F(R', \Theta) = - \frac{k_1(R')\, \Theta}{1+2k_2(R')\,\Theta + k_3(R')\,\Theta^2}.
    \label{eq:Fs}
\end{equation}
The coefficients $k_1, k_2, k_3$ are positive analytical functions of the scaled radius, and their 
precise definition can be found, for instance, in \citet{1998A&A...335.1093V,vokrouhlicky_1999}. 
It is worth noting that the seasonal component always produces an inward migration, while
the direction of migration for the diurnal component depends on the obliquity $\gamma$. 

\end{document}